# Pockels effect induced strong Kerr nonlinearity in a lithium niobate waveguide

Haoran Li,[1] Fei Huang,[1] Jingyan Guo,[1] He Gao,[1] Hanwen Li,[1] Zhile Wu,[1] Xinmin Yao,[1] Zhengyuan Bao,[1] Huan Li,[1,2,3] Yaocheng Shi,[1,2,3] Zejie Yu,[1,2,3,*] and Daoxin Dai[1,2,3]

[1]State Key Laboratory of Extreme Photonics and Instrumentation, College of Optical Science and Engineering, Zhejiang University, Hangzhou 310058, China

[2]ZJU-Hangzhou Global Scientific and Technological Innovation Center, Zhejiang University, Hangzhou 311215, China

[3]Jiaxing Key Laboratory of Photonic Sensing & Intelligent Imaging, Intelligent Optics & Photonics Research Center, Jiaxing Research Institute, Zhejiang University, Jiaxing 314000, China

*zjyu@zju.edu.cn

**Abstract:** The utilization of Kerr nonlinearity in lithium niobate has been extensively investigated over the years. Nevertheless, the practical implementation of Kerr nonlinearity in waveguides has been constrained by the material's inherently low third-order nonlinear coefficients. Here, we present a significant advancement by demonstrating Pockels effect-induced strong Kerr nonlinearity in a periodically poled thin-film lithium niobate waveguide. Both effective four-wave mixing (FWM) and cascaded effective FWM processes are experimentally observed. The induced FWM process achieves a remarkable maximum output power of -8.5 dBm, spanning a wavelength spectrum of over 116.8 nm. Analysis reveals that the induced effective Kerr nonlinearity exhibits a substantial effective nonlinear refractive index $n_{2,\text{eff}}$ as $2.9 \times 10^{-15}$ m$^2$/W, corresponding to an effective nonlinear refractive index enhancement factor of $1.6 \times 10^4$ relative to the intrinsic value. Moreover, a wavelength-converting experiment demonstrates a flat optic-to-optic response over a broadband radiofrequency spectrum, confirming that signal integrity is well preserved after on-chip effective FWM conversion. Therefore, the demonstrated efficient and broadband Pockels effect induced effective Kerr nonlinearity pave the way for novel applications in diverse fields, including spectroscopy, parametric amplification, quantum correlation studies, and wavelength conversion technologies.

**Key words:** lithium niobate, four-wave mixing, Pockels effect

## 1. INTRODUCTION

Thin film lithium niobate (TFLN) has been widely explored for versatile integrated photonic devices owing to its extraordinary optical properties[1–5] such as low loss, a broadband transparency window (400 nm to 5 μm), and large linear electro-optic (EO) coefficients ($r_{33}$ = 32.6 pm/V). In addition, the strong confinement of optical fields enhances the light-matter interaction strength, significantly reducing the pump power and device footprint needed compared to bulky lithium niobate[6–8]. Kerr nonlinearity, a third-order nonlinear property that exists in almost all types of materials, plays an important role in Kerr frequency comb generation[2,9,10], spectroscopy[11], parametric amplification[12], quantum correlation[13], and wavelength conversion[14]. Utilizing Kerr nonlinearity in TFLN has long been pursued[15]. In 2019, C. Wang et al. first demonstrated on-chip Kerr frequency comb generation in a dispersion-engineered TFLN microring cavity using third-order nonlinearity[16]. However, the Kerr nonlinearity strength for lithium niobate ($n_2$ = 1.8×10$^{-19}$ m$^2$/W @1550 nm) is relatively small compared with that of silicon ($n_2$ = 5.0×10$^{-18}$ m$^2$/W @1550 nm), restricting many applications requiring strong Kerr nonlinearity[17].

Leveraging the substantial Pockels effect inherent in lithium niobate, the implementation of cascaded quadratic nonlinear processes emerges as a promising approach for realizing effective Kerr nonlinearity in this material system. In 2017, Shijie et al. presented a third-harmonic generation (THG) process using cascaded second-harmonic generation (SHG) and sum-frequency generation (SFG) processes in a TFLN microdisk cavity[18]. In 2019, Shijie L. demonstrated effective four-wave-mixing (FWM) in TFLN microdisk cavity using cascaded SHG and difference-frequency generation (DFG) processes[19]. Apart from TFLN, aluminum nitride microring[20] and thin-film lithium tantalate microdisk[21] cavities have also been demonstrated to achieve effective Kerr nonlinearity using cascaded quadratic processes. However, it is crucial to note that these demonstrated implementations predominantly rely on microcavity architectures, which impose significant constraints on both pump power and conversion bandwidth. The input pump power in such systems is restricted because light enhancement in a microcavity can induce detrimental thermal and photorefractive effects, substantially disrupting the delicate phase-matching conditions within the cavity[21]. Furthermore, the strong dispersion characteristics of the interacting waves in the microcavity fundamentally limit the achievable conversion bandwidth, presenting a critical challenge for practical applications.

Here, we present a significant advancement by demonstrating Pockels-effect-induced strong Kerr nonlinearity in a periodically poled TFLN waveguide. The demonstrated periodically poled TFLN waveguide realizes a strong SHG with conversion efficiency up to 1896.81 %W$^{-1}$, enabling effective FWM and cascaded effective FWM based on cascaded Pockels effects without the assistance of a microcavity to significantly enhance light intensity. Therefore, limitations of pump power and bandwidth for effective Kerr nonlinearity can be overcome. The induced effective FWM demonstrates exceptional performance metrics, achieving a remarkable wavelength spectrum exceeding 116.8 nm. Through rigorous comparative analysis, we observe an effective nonlinear refractive index $n_{2,eff}$ is evaluated as 2.9×10$^{-15}$ m$^2$/W, which translates to an effective nonlinear refractive index enhancement factor of 1.6×10$^4$ relative to the intrinsic value. Furthermore, wavelength conversion experiments reveal a flat optical-to-optical (OO) response across a broad radiofrequency spectrum, unequivocally demonstrating the preservation of signal integrity following on-chip effective FWM conversion. These results not only establish a new paradigm for enhancing effective Kerr nonlinearity in the TFLN platform but also open exciting avenues for diverse applications in nonlinear photonics, quantum optics, and integrated photonic circuits.

## 2. DEVICE DESIGN AND FABRICATION

Figure 1a shows a schematic illustration of the proposed induced effective Kerr nonlinearity device based on a periodically poled TFLN waveguide. The inset shows a cross-sectional view of the TFLN waveguide with a top width $w$ of 2 μm, waveguide thickness $t$ of 600 nm, and etch depth $h$ of 300 nm. The sidewall angle of the waveguide was approximately 60° due to the isotropic etching process of the TFLN waveguide.

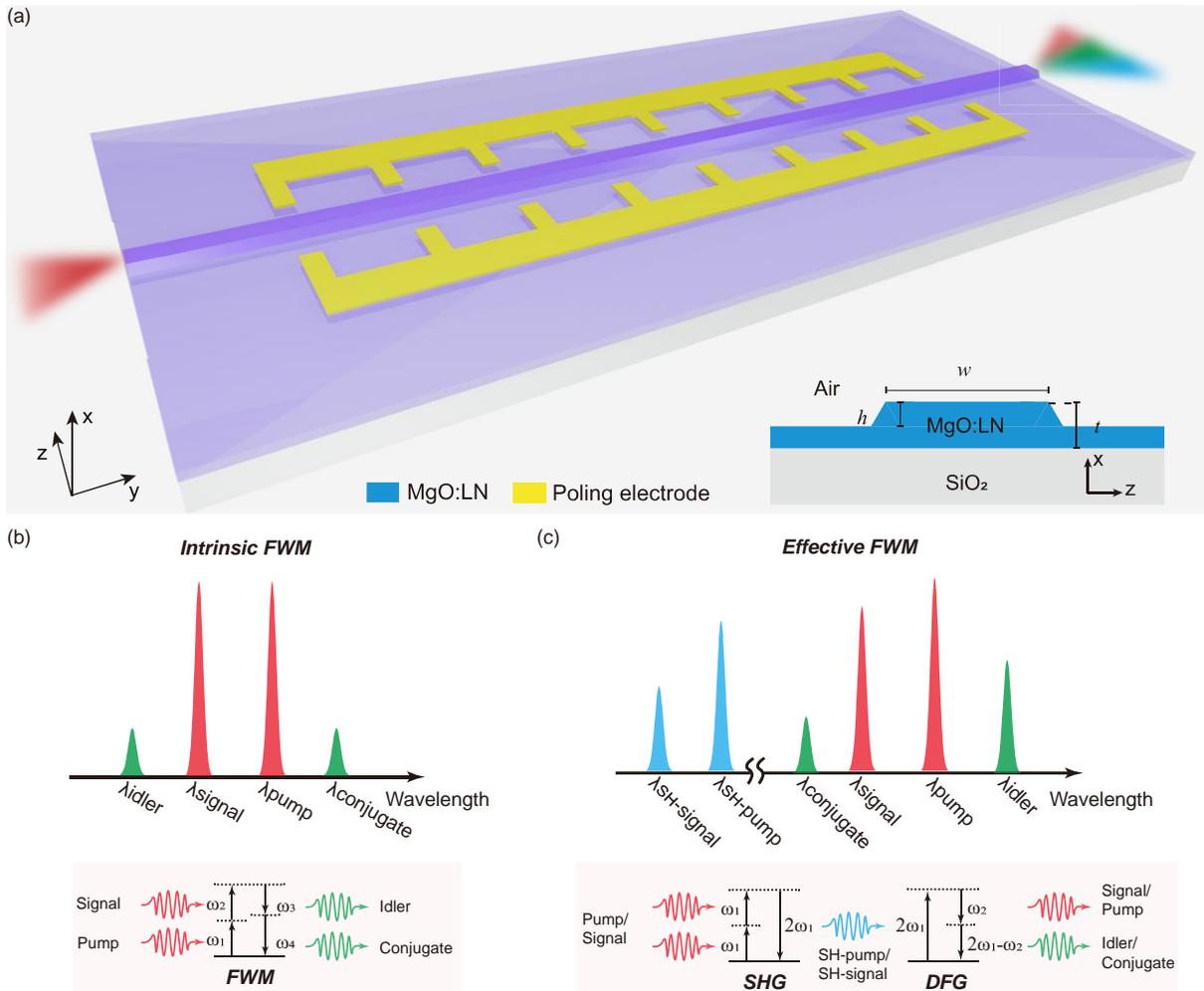

Figure 1. (a) Schematic illustration and cross-section view of the effective Kerr nonlinearity device based on a periodically poled TFLN waveguide. (b) Illustration of intrinsic Kerr nonlinearity in a TFLN waveguide. (c) Illustration of effective Kerr nonlinearity in a periodically poled TFLN waveguide based on cascaded SHG and DFG processes.

Figure 1b demonstrates the intrinsic Kerr nonlinear process in the waveguide. When energy conservation and phase-matching condition are both satisfied, idler and conjugate waves can be converted from injected pump and signal waves using the Kerr nonlinearity of lithium niobate. Figure 1c depicts the main quadratic processes in the periodically poled TFLN waveguide. The effective FWM originates the cascading of two quadratic (SHG and DFG) processes, where the pump wave can be converted to a second harmonic (SH) wave through the SHG process, and the SH wave converts the signal wave into an idler wave through a DFG process. A conjugated wave is generated in a similar way to the idler wave, where the SH wave from the signal wave interacts with the pump wave through a DFG process. The cascaded quadratic processes can be seen as an equivalent Kerr process where pump and signal waves can be converted to idler and conjugate waves near the pump and signal wavelengths. By using a non-depletion regime and ignoring the phase mismatches between the interacting waves, the effective Kerr nonlinearity induced nonlinear refractive index $n_{2,eff}$ can take the following form (see in Supplementary Information S1)

$$n_{2,eff} = \frac{cA_{eff}}{2\omega z\sqrt{P_p P_s}} \text{arcsinh}(\sqrt{\frac{CE \cdot P_s}{P_i(0)}}) \tag{1}$$

where $CE$ and $A_{eff}$ refer to the conversion efficiency and effective mode area of effective FWM in the waveguide. And $\omega$ denotes average angular frequency of pump, signal and idler wave while $z$ denotes the propagation length in the waveguide. Using a low-loss waveguide with strong second-order nonlinearity, the effective nonlinear refractive index is expected to be larger than the intrinsic nonlinear refractive index $n_2$ induced by intrinsic Kerr nonlinearity. The nonlinear refractive index enhancement factor can be defined as:

$$\Gamma = \frac{n_{2,eff}}{n_2} \tag{2}$$

The conversion efficiency (CE) of effective FWM can be evaluated as:

$$CE = \frac{P_i(L)}{P_s(0)} \cdot 100\% \tag{3}$$

where $L$ refers to the length of the periodically poled TFLN waveguide, $P_i$ and $P_s$ represent the power of the idler wave at the output and signal wave at the input of the periodically-poled TFLN waveguide.

Figure 2a illustrates the effective refractive index curves of the fundamental TE mode across the wavelength spectrum of 1520–1620 nm (blue), along with its corresponding SH wave (red) in the designed TFLN waveguide. To satisfy the phase-matching condition, periodic poling is essential to address the effective refractive index difference between the pump wave and its associated SH wave. The inset of Figure 2a shows the modal profile of the fundamental TE mode. Figure 2b displays the simulated phase mismatches (red) in the absence of periodic poling, alongside the corresponding poling periods (blue) required for SHG to compensate for phase mismatches. Similarly, Figure 2c presents the simulated phase mismatches (red) without periodic poling and the corresponding poling periods (blue) for DFG to achieve phase matching. Notably, the phase mismatches in DFG exhibit minimal wavelength dependence compared to those in SHG. This is primarily due to the idler wave in DFG experiencing an opposite refractive index change relative to the signal wave as the signal wavelength is varied. Figure 2d presents the simulated normalized conversion efficiency (NCE) of second harmonic generation (SHG) for the pump wave in periodically poled TFLN waveguides with lengths of 5 mm and 10 mm. The SHG bandwidth is notably narrow due to the wavelength-dependent phase mismatch between the pump and second harmonic waves, as periodic poling can only compensate for phase mismatch at a specific wavelength. While longer waveguide lengths yield higher SHG conversion efficiency, they also result in a narrower conversion bandwidth due to the accumulation of phase mismatch over the extended propagation distance. Figure 2e illustrates the simulated NCE of DFG between a fixed SH wave generated from a pump wave at 1560 nm and a signal wave with varying wavelengths. The DFG process exhibits a broad working bandwidth exceeding 295 nm, demonstrating its versatility across a wide spectral range. Figure 2f plots the NCE of effective FWM for waveguide lengths of 5 mm (blue) and 10 mm (red). It is evident that effective FWM benefits from a broadband bandwidth similar to DFG, as the SHG-related pump wave operates at a fixed wavelength. Consequently, the bandwidth of the effective FWM process is primarily determined by the DFG process, enabling efficient nonlinear interactions over a wide spectral range[22,23].

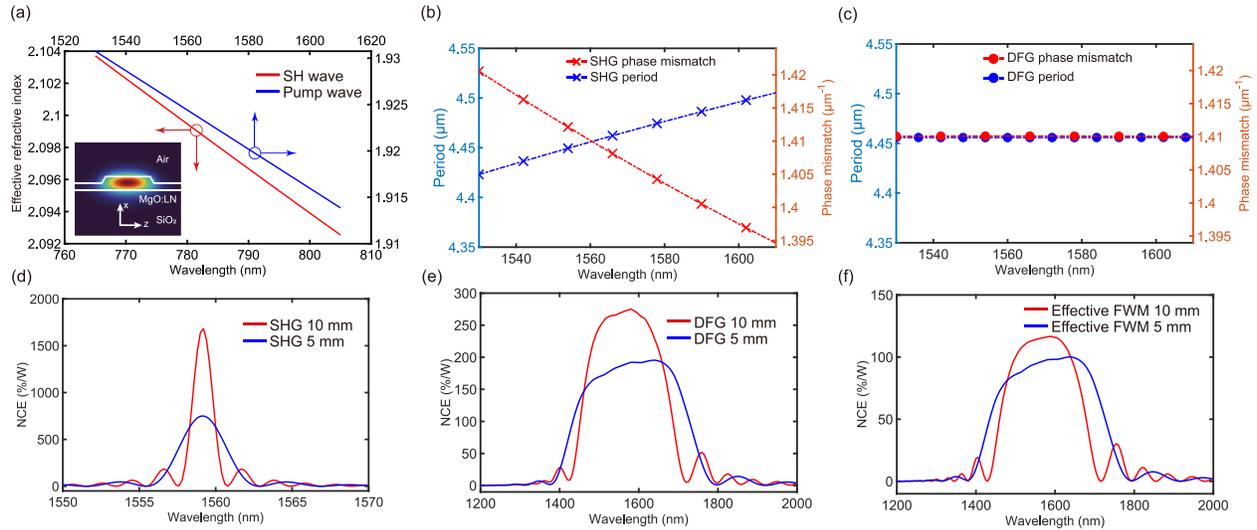

Fig. 2. (a) Simulated effective refractive index curves of the fundamental TE mode at a wavelength spectrum of 1520–1620 nm and its corresponding SH wave in the designed TFLN waveguide. The inset shows the modal profile of the fundamental TE mode. (b) Simulated phase mismatches and required poling periods of SHG at different wavelengths. (b) Simulated phase mismatches and required poling periods of DFG at different wavelengths. (d) Simulated NCE of SHG process in 10-mm (red) and 5-mm (blue) waveguides. (e) Simulated NCE of DFG process in 10-mm (red) and 5-mm (blue) waveguides. (f) Simulated NCE of effective FWM process in 10-mm (red) and 5-mm (blue) waveguides.

## 3. DEVICE CHARACTERIZATION

The device was fabricated on a 600-nm-thick x-cut 5% MgO-doped TFLN layer, supported by a 2-μm silica buffer and a 525-μm-thick silicon handle substrate. Figure 3a displays an optical microscope image of the fabricated TFLN waveguides, while Figure 3b presents a piezoelectric force microscope (PFM) image, clearly revealing the uniform periodic domain inversion achieved in the TFLN waveguide. Figure 3c shows a scanning electron microscope (SEM) image of one facet of the chip. To maximize the light coupling efficiency between the lensed fiber and the waveguide, the waveguide width at the chip edge was tapered to 4 μm. The measured coupling loss between the lensed fiber and the waveguide is approximately 6.5 dB. To evaluate the performance of the fabricated periodically poled TFLN waveguide, its SHG response was characterized. The measured and simulated normalized conversion efficiencies (NCEs) of SHG are illustrated in Figure 3d. When the pump wavelength is tuned to 1551.1 nm, a peak NCE of 1896.81 %W$^{-1}$ and a maximum on-chip power of 7.0 dBm are achieved at a temperature of 30 °C. Figure 3e presents the measured SHG wavelength as a function of temperature, revealing a temperature-dependent shifting coefficient of 214 pm/°C. Figure 3f shows the NCE of SHG as a function of pump power, demonstrating a quadratic dependence on the pump power, as expected for a second-order nonlinear process.

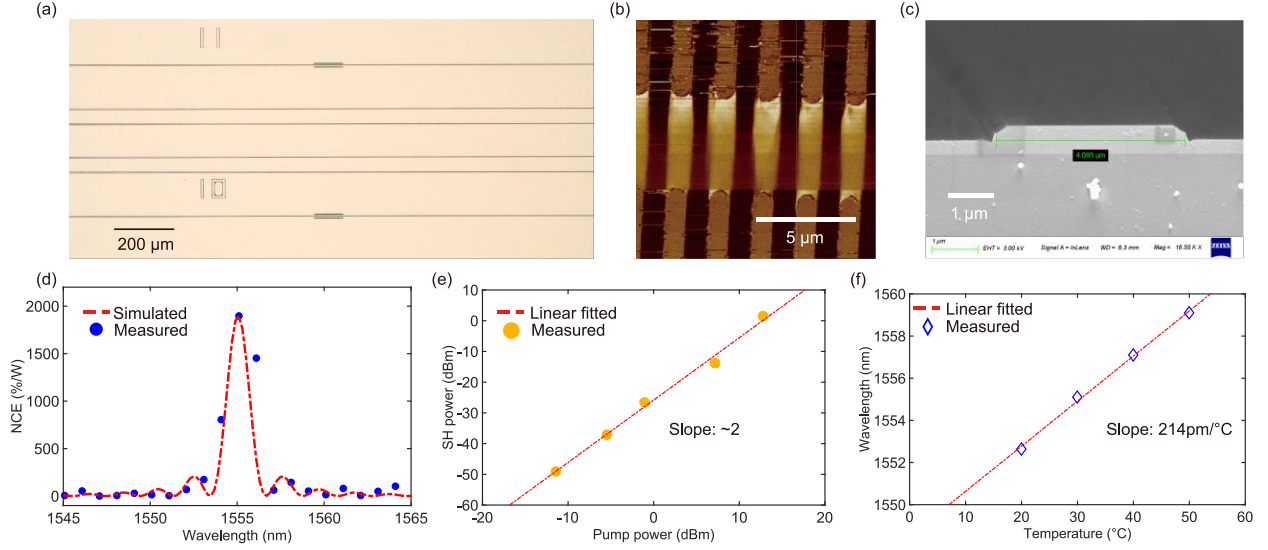

Fig. 3. (a) Optical microscope image of the fabricated periodically poled TFLN waveguide. (b) PFM image of the fabricated periodically poled TFLN waveguide. (c) SEM image of polished TFLN waveguide facet. (d) Measured and simulated NCE of SHG in the fabricated periodically poled TFLN waveguide. (e) The output power of the SH wave as a function of the pump wave power. (f) The SHG wavelength as a function of temperature.

The experimental setup for characterizing the effective FWM phenomenon is depicted in Figure 4a, where pump and signal waves are sent into the TFLN waveguides simultaneously. The fabricated chip under the experimental setup is shown in Figure 4b. Figure 4c presents the measured optical spectrum under a pump wave with a power of 12.4 dBm and a wavelength of 1560 nm, as well as a signal wave with a power of 16.0 dBm and a wavelength of 1533 nm, where the temperature of the chip is tuned to 50 °C. An idler power of −8.5 dBm and CE of −24.5 dB is achieved. Since the $A_{eff}$, $\omega$, $CE$, $z$, $P_p$, and $P_s$ are evaluated as $1.15\times10^{-12}$ m², $2\pi\times1.93\times10^{14}$ Hz, −24.5 dB, 1 cm, 0.0174 W, and 0.0398 W, respectively, the effective nonlinear refractive index $n_{2,eff}$ can be calculated as $2.9\times10^{-15}$ m²/W under the assumption that $P_i(0)$ is $10^{-8}$ W. Moreover, considering the intrinsic nonlinear refractive index as $1.8\times10^{-19}$ m²/W, the nonlinear refractive index enhancement $\Gamma$ can be derived as $1.6\times10^4$. Unlike intrinsic FWM, several waves resulting from the SHG and SFG of the pump and signal waves appear at the SH spectrum besides the emergence of idler and conjugate waves. Figure 4f shows the measured effective idler wave power under different pump and signal wave powers. The measured slopes of the power of the idler wave with the power of the pump and signal waves are approximately 1 and 2, respectively. In other words, the power of the idler wave increases linearly with the power of the signal wave but quadratically with the power of the pump wave. It can be attributed to the intrinsic conversion relationships of effective FWM, where the idler wave originates from the DFG process between the SH wave converted from the pump wave and the signal wave. According to the evolution law of DFG, the idler wave power increases linearly with the power of the SH wave converted from the pump wave and signal wave. However, the power of the SH wave increases quadratically with the power of the pump wave because of the evolution law of SHG. The recorded optical spectra are shown in Figure 4d when sweeping the signal wave wavelength from 1530.0 nm to 1606.5 nm. It can be seen that the idler and conjugate waves were effectively generated during the whole sweep, and their wavelengths were highly consistent with the predictions of the cascaded SHG and DFG processes, concluding that the effective FWM wave had a generation bandwidth of over 116.8 nm. Figure 4e shows the measured CE (dots) of effective FWM as a function of signal wave wavelength in the periodically poled TFLN waveguide under the condition that the wavelength of the pump wave is fixed at 1560.0 nm and the temperature of the chip is fixed at 50 °C. The measured CE of effective FWM shows a broadband performance, agreeing well with the simulated spectrum (line). An identical waveguide without periodic poling fabricated on the same chip has been characterized to show that the idler and conjugated waves originated from effective FWM instead of intrinsic FWM. Figure 4g presents the measured optical spectrum after transmitting through the waveguide without periodic poling under identical pump and signal wave conditions. An extremely weak intensity of intrinsic idler and conjugate waves is observed, and no SH or SF waves are detected. These results demonstrate a significant enhancement of effective Kerr nonlinearity through the application of cascaded quadratic processes in the periodically poled TFLN waveguide.

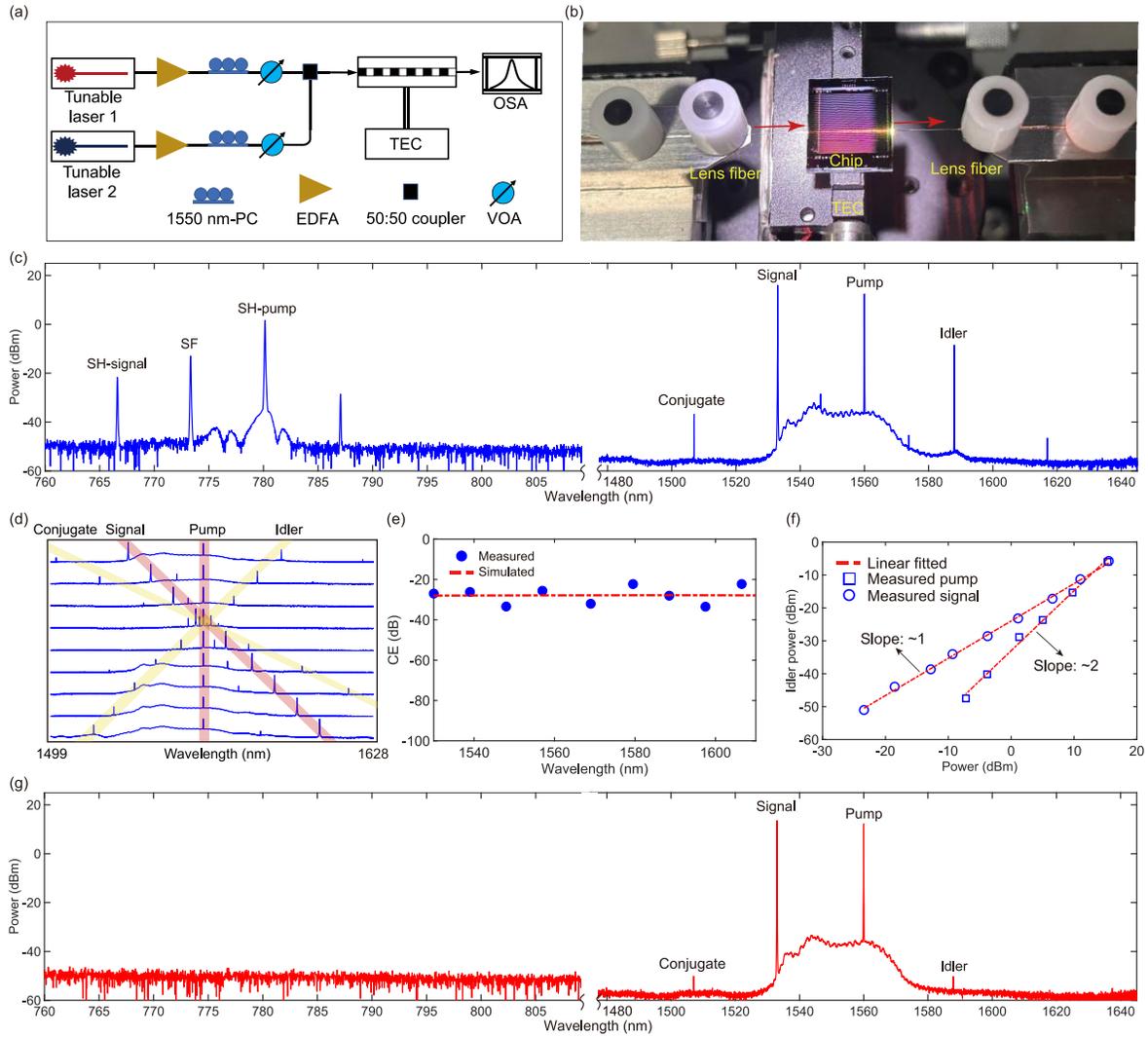

Figure 4. (a) Experimental setup for effective FWM characterization. TEC, thermoelectric cooler. PC, polarization controller. EDFA, erbium-doped fiber amplifier. OSA, optical spectrum analyzer. VOA, variable optical attenuator. (b) Captured image of the chip in the experimental setup. (c) Measured optical spectrum under a pump and signal waves with a wavelength of 1560.0 nm and 1533.0 nm, respectively. (d) Measured on-chip idler power under different powers of pump and signal waves. (e) Measured optical spectra in the pump and signal regions in the periodically poled TFLN waveguide during signal wavelength sweep. (f) Measured and simulated CEs of effective FWM during signal wavelength sweep. (g) Measured optical spectrum in a TFLN waveguide without poling when the pump and signal were fixed at 1560.0 nm and 1533.0 nm, respectively.

The quasi-phase matching conditions for both the signal and pump waves are simultaneously satisfied to achieve efficient SHG when the signal wavelength is tuned close to that of the pump wave. When both the pump and signal waves possess sufficient power, cascaded effective FWM phenomena can be observed. With the pump and signal wavelengths fixed at 1560.0 nm and 1558.5 nm, respectively, Figure 5 illustrates the evolution of the measured optical spectrum at varying on-chip input powers. As the input power increases, the effective FWM progressively develops into cascaded effective FWM. At an on-chip input power of 13 dBm for both pump and signal waves, a distinctive comb-like spectrum emerges, featuring 11 wavelengths around 1560 nm and 6 wavelengths in the second-harmonic spectrum. This observation provides compelling evidence for the occurrence of highly efficient and strong effective Kerr nonlinearity. It is anticipated that further enhancement of the on-chip input powers for both pump and signal waves would lead to the generation of additional cascaded effective FWM waves, potentially expanding the spectral range and complexity of the nonlinear interactions.

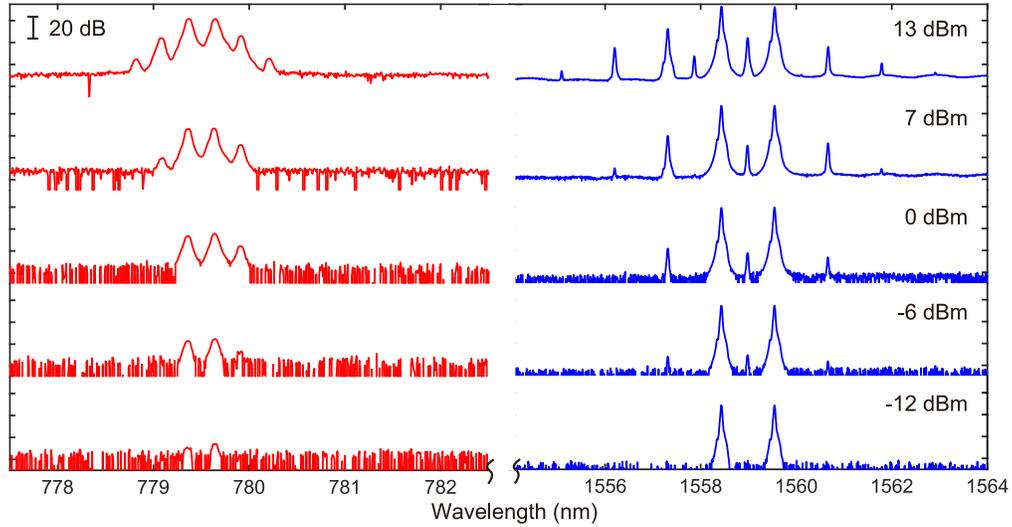

Figure 5. Measured optical spectra at varying on-chip input powers, the effective FWM progressively develops into cascaded effective FWM.

To further demonstrate the application of effective FWM generation in the periodically poled TFLN waveguide, a transmission experiment was conducted to show the ability to change signal wavelength without losing the information it carried. Figure 6a demonstrates the experimental setup for the transmission experiment. An EO response of signal wave $S_{21,signal}$ can be obtained by modulating and measuring the signal wave using a light component analyzer (LCA). An EO response of the idler wave $S_{21,idler}$ can be obtained by modulating the signal wave but measuring the response of the idler wave, which characterizes the signal integrity maintenance after effective FWM. An optic-to-optic (OO) response $S_{21,FWM}$, characterizing the signal integrity maintenance in the wavelength conversion system, can be calculated by differencing the two EO responses $S_{21,idler}$ and $S_{21,signal}$

$$S_{21,FWM} = S_{21,idler} - S_{21,signal} \qquad (4)$$

The normalized OO response of DFG $S_{21,\,NORM}$ can be evaluated as

$$S_{21,NORM} = S_{21,FWM} - S_{21,FWM}(1\ GHz) \qquad (5)$$

In the transmission experiment, the signal and pump were fixed at 1570.0 nm and 1559.5 nm, respectively. After amplification by EDFAs, the signal and pump waves had on-chip input powers of 7.8 dBm and 14.5 dBm, respectively. The measured EO responses of signal and idler waves are shown in Figure 6b. The corresponding normalized OO response plotted in Figure 6c has a flat and broadband spectrum, indicating extraordinary signal integrity maintenance after wavelength conversion in effective FWM. Figure 6d presents the measured optical spectrum obtained from the OSA, demonstrating complete suppression of the signal wave through band-pass filtering. This confirms that the EO response recorded by the LCA originates from the idler wave. Figure 6e provides a detailed view of the idler wave spectrum, observed sidebands verify successful transfer of the modulation from the signal to the idler waves through the effective FWM process.

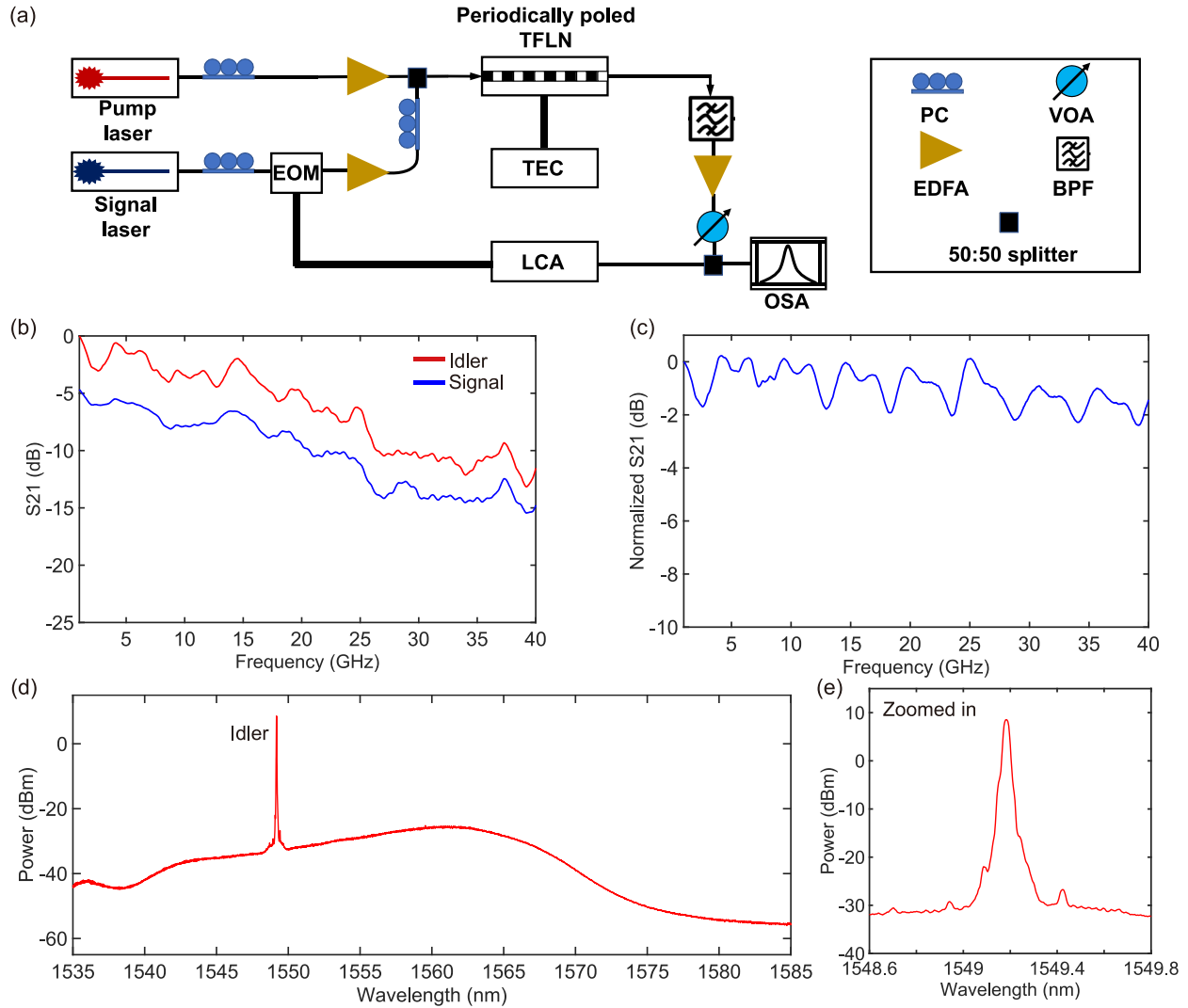

Figure 6. (a) Experiment setup for transmission experiment. (b) Measured EO response of the signal and idler waves in the system. (c) Calculated normalized OO response of the effective FWM process. (d) Measured optical spectrum obtained from the OSA. (e) Zoomed-in idler wave spectrum.

Table 1. Performance comparison of Pockels induced effective Kerr nonlinearity in integrated devices

| Platform | Type | Pump power (mW) | Converted power (mW) | Bandwidth (nm) |
| --- | --- | --- | --- | --- |
| Lithium niobate microdisk [18] | THG | 20 | 0.00038[a] | Single wavelength |
| Lithium niobate microdisk [19] | eFWM | 5 | 0.005[a] | Single wavelength |
| Aluminum nitride microring [20] | eFWM | ~ 0.006[a] | ~0.0000016[a] | Single wavelength |
| Lithium tantalite microdisk [21] | THG | 24 | 0.00002 | Single wavelength |
| This work (lithium niobate waveguide) | eFWM | 57.19 | 0.14 | >116 |

[a]Calculated from relevant data demonstrated in the paper.

We present a comprehensive performance comparison of our device's effective Kerr nonlinearity with other integrated platforms in Table 1. Notably, our system demonstrates a significant enhancement in conversion efficiency, with the absolute power of the converted wave exceeding that of existing implementations by several orders of magnitude. Furthermore, the waveguide-based architecture exhibits an exceptional broadband response, effectively overcoming the bandwidth constraints typically associated with cavity-based devices. This broadband capability represents a substantial advancement in nonlinear photonic integration, offering unprecedented operational flexibility for wavelength conversion applications.

## 4. DISCUSSION AND CONCLUSION

In conclusion, we demonstrated a device generating effective Kerr nonlinearity induced by Pockels effects in a periodically poled TFLN waveguide. Empowered by high-efficiency quadratic processes of SHG and DFG in the waveguide, both effective and cascaded effective FWMs are experimentally observed without any cavity structures. The induced FWM process achieves a remarkable maximum output power of -8.5 dBm, spanning a wavelength spectrum of over 116.8 nm, breaking the limitations of pump power and complex dispersion issues encountered by cavities. Besides, the effective nonlinear refractive index $n_{2,eff}$ is evaluated as $2.9 \times 10^{-15}$ m$^2$/W, which translates to an effective nonlinear refractive index enhancement factor of $1.6 \times 10^4$ relative to the intrinsic value. Moreover, the flatness OO response of effective FWM generation showed excellent signal integrity maintenance in the wavelength conversion process. The strong effective FWM induced by the quadratic process paved the way for using the effective Kerr nonlinearity in the TFLN platform, which may find applications in spectroscopy, parametric amplification, quantum correlation and wavelength conversion.

## METHODS

**Fabrication.** The precise thickness of the TFLN layer was determined using an optical thickness measurement instrument based on reflectance analysis. A two-dimensional motorized stage was positioned beneath the TFLN chip to accurately map the thickness distribution across the sample. Poling electrodes with customized periods to satisfy the phase-matching condition with the non-uniform thickness profile along the waveguide were patterned using electron-beam lithography (EBL). Subsequently, a multilayer stack of 100 nm Al$_2$O$_3$, 10 nm Ti, and 150 nm Au was deposited through sequential electron-beam evaporation and lift-off processes. Periodic poling was achieved by a series of short high-voltage pulses through the fabricated electrodes. Following this, another EBL step was employed to define the waveguide pattern, which was then transferred into the TFLN layer via Ar+ plasma dry etching, resulting in waveguides with a sidewall angle θ of approximately 60°. Finally, the chip was diced and polished to facilitate efficient light coupling between the lensed fibers and the waveguides. The periodic ferroelectric domain inversion in the poled waveguide was characterized using PFM, confirming the successful fabrication of the device.

**Effective FWM performance characterization.** Pump and signal waves were generated by two tunable continuous-wave (CW) laser sources and subsequently amplified using two erbium-doped fiber amplifiers (EDFAs). A fiber-based 50:50 broadband coupler was employed to combine the signal and pump waves into a single fiber. To ensure the utilization of the maximum second-order nonlinear coefficient ($d_{33}$) of lithium niobate, two polarization controllers (PCs) were integrated into the setup to align the pump and signal waves into the fundamental TE modes. The on-chip power levels of all waves were precisely extracted from the spectral data obtained using an optical spectrum analyzer (OSA).

**OO response characterization.** The signal wave generated by a tunable CW laser operating at 1570.0 nm wavelength is first adjusted through a PC before being directed into an EO modulator. Within the modulator, high-speed data from an LCA is encoded onto the optical carrier. The modulated signal wave, along with a pump wave, was independently polarization-controlled using two separate PCs. These optical signals were then combined using a 50:50 fiber-based power splitter and efficiently coupled into a periodically-poled TFLN waveguide through a pair of lensed fibers. The output light was filtered using a band-pass optical filter to isolate the idler wave. To achieve optimal measurement conditions, the idler wave was subsequently amplified using an EDFA and precisely attenuated to the desired power level through a variable optical attenuator. Finally, the processed idler wave was split using another 50:50 fiber-based splitter, with one path directed to an OSA for spectral characterization and the other to the LCA for signal analysis.

**Funding.** National Natural Science Foundation of China (62450079) and the Zhejiang Provincial Natural Science Foundation of China (LDT23F04012F05).

**Disclosures.** The authors declare no conflicts of interest.

**Data availability.** The data underlying the results presented in this paper are not publicly available at this time but may be obtained from the authors upon reasonable request.

# Supplementary Information for

# Pockels effect induced strong Kerr nonlinearity in a lithium niobate waveguide

Haoran Li,[1] Fei Huang,[1] Jingyan Guo,[1] He Gao,[1] Hanwen Li,[1] Zhile Wu,[1] Xinmin Yao,[1] Zhengyuan Bao,[1] Huan Li,[1,2,3] Yaocheng Shi,[1,2,3] Zejie Yu,[1,2,3,*] and Daoxin Dai[1,2,3]

[1]State Key Laboratory of Extreme Photonics and Instrumentation, College of Optical Science and Engineering, Zhejiang University, Hangzhou 310058, China

[2]ZJU-Hangzhou Global Scientific and Technological Innovation Center, Zhejiang University, Hangzhou 311215, China

[3]Jiaxing Key Laboratory of Photonic Sensing & Intelligent Imaging, Intelligent Optics & Photonics Research Center, Jiaxing Research Institute, Zhejiang University, Jiaxing 314000, China

*zjyu@zju.edu.cn

In a cascaded SHG and DFG process, the pump and signal waves are simultaneously injected into a periodically poled TFLN waveguide. If the energy conservation and phase-matching conditions for the SHG and DFG processes are both satisfied, the SHG and DFG processes can simultaneously take place in the thin-film PPLN waveguide. The power of the pump wave will flow to its SH wave through the SHG process and eventually flow to the idler wave through the DFG process. These two cascading processes can be seen an effective FWM process. Similarly, the power of the signal wave can flow to the conjugate wave. Taking the idler generation process for example, the coupled-wave equations describing the effective FWM process take the following form:

$$\frac{\partial A_p}{\partial z} = -\frac{\alpha_p}{2} A_p - i\kappa A_{SH} A_p^* e^{-i\Delta\beta_{SHG} z} \tag{S1}$$

$$\frac{\partial A_{SH}}{\partial z} = -\frac{\alpha_{SH}}{2} A_{SH} - i\kappa A_p^2 e^{+i\Delta\beta_{SHG} z} - i2\kappa A_s A_i e^{i\Delta\beta_{DFG} z} \tag{S2}$$

$$\frac{\partial A_s}{\partial z} = -\frac{\alpha_s}{2} A_s - i\kappa A_i^* A_{SH} e^{-i\Delta\beta_{DFG} z} \tag{S3}$$

$$\frac{\partial A_i}{\partial z} = -\frac{\alpha_i}{2} A_i - i\kappa A_s^* A_{SH} e^{-i\Delta\beta_{DFG} z} \tag{S4}$$

where $\alpha_j$ ($j = p, s, i,$ and $SH$) refers to the propagation loss at the respective wavelength in the waveguide, $A_j/\kappa_j$ ($j = p, s, i,$ and $SH$) refers to the amplitude/coupling coefficient of the pump, signal, idler, and SH waves, respectively, and $z$ refers to the propagation distance in the periodically poled TFLN waveguide. Considering the wavelengths of pump, signal, and idler waves are close, the coupling coefficient in the waveguide can take the following form:

$$\kappa = \frac{d_{eff}\omega}{cn} \cdot \frac{\iint \epsilon_s(x,y)\epsilon_i(x,y)\epsilon_p(x,y)dxdy}{\iint \epsilon_a^2(x,y)dxdy} \tag{S5}$$

where $d_{eff}$ denotes the effective nonlinear coefficient of the DFG process, $\varepsilon_j(x,y)$ ($j = p, s, i, SH,$ and $a$) denotes the cross-sectional modal distribution of pump wave, signal wave, idler wave, SH wave, and their average modal distribution in the waveguide, $\omega$ and $n$ refer to the average angular frequency and effective refractive index, respectively. If the largest second-order nonlinear coefficient of lithium niobate $d_{33}$ is used, the effective nonlinear coefficient can be written as [S1]

$$d_{eff} = d_{33}\frac{2sin(K\pi D)}{K\pi} \tag{S6}$$

where $D$ is the duty cycle of periodical poling in a TFLN waveguide and $K$ is the order of the Fourier coefficients of the periodical poling profile along the propagation orientation.

Moreover, $\Delta\beta_{SHG}$ and $\Delta\beta_{DFG}$ refer to the effective propagation constant mismatches of the interacting waves of SHG and DFG, respectively. When the effective propagation constant mismatch is zero, the phase-matching condition is satisfied, and an efficient SHG and DFG process will occur. In a periodically poled TFLN waveguide, effective propagation constant mismatches can be expressed in the following form:

$$\Delta\beta_{SHG} = 2\pi\left(\frac{n_{eff,SH}}{\lambda_p} - 2\frac{n_{eff,p}}{\lambda_p} - \frac{m}{\Lambda}\right) \tag{S7}$$

$$\Delta\beta_{DFG} = 2\pi\left(\frac{n_{eff,p}}{\lambda_p} - \frac{n_{eff,s}}{\lambda_s} - \frac{n_{eff,i}}{\lambda_i} - \frac{m}{\Lambda}\right) \tag{S8}$$

$n_{eff,j}$ and $\lambda_j$ ($j = p, s, i,$ and $SH$) refer to the effective refractive indices and vacuum wavelength of the pump, signal, idler, and SH waves, respectively. $\Lambda$ refers to the poling period in the thin-film PPLN waveguide, and $m$ is the index of the phase-matching order.

To gain a deeper understanding of effective FWM, an approximate analytical solution is derived under certain simplifying conditions. First, we assume that propagation loss and group velocity mismatch can be neglected. By invoking the non-depletion regime and perfect phase-matching conditions, the SHG and DFG processes become decoupled. Consequently, the output power of the idler wave can be expressed as:

$$P_i \approx \frac{1}{4}\frac{2\pi^2 d_{eff}^2}{\varepsilon_0 c n_{eff,SH} n_{eff,p}^2 \lambda_{SH}^2 S_{SHG,eff}} \frac{8\pi^2 d_{eff}^2}{\varepsilon_0 c n_{eff,i} n_{eff,s} n_{eff,p} \lambda_i \lambda_s S_{DFG,eff}} L^4 P_{p,0}^2 P_{s,0} \tag{S9}$$

where $P_{p,0}$ and $P_{s,0}$ denote the input power of the pump and signal waves, $S_{SHG,eff}$ and $S_{DFG,eff}$, denoting the effective cross section in the SHG and DFG processes, can be calculated using the normalized overlap integral of the modal distributions of interacting waves:

$$S_{SHG,eff} = \frac{(\iint \epsilon_p^2(x,y)dxdy)^2 \iint \epsilon_{SH}^2(x,y)dxdy}{(\iint \epsilon_p^2(x,y)\epsilon_{SH}(x,y)dxdy)^2} \tag{S10}$$

$$S_{DFG,eff} = \frac{\iint \epsilon_{SH}^2(x,y)dxdy \iint \epsilon_s^2(x,y)dxdy \iint \epsilon_i^2(x,y)dxdy}{(\iint \epsilon_{SH}(x,y)\epsilon_s(x,y)\epsilon_i(x,y)dxdy)^2} \tag{S11}$$

The approximate conversion efficiency of this cascading process can then be expressed as follows:

$$\eta = \frac{P_i}{P_{s,0}} \approx \frac{1}{4}\frac{2\pi^2 d_{eff}^2}{\varepsilon_0 c n_{eff,SH} n_{eff,p}^2 \lambda_{SH}^2 S_{SHG,eff}} \frac{8\pi^2 d_{eff}^2}{\varepsilon_0 c n_{eff,i} n_{eff,s} n_{eff,p} \lambda_i \lambda_s S_{DFG,eff}} L^4 P_{p,0}^2 \tag{S12}$$

In an intrinsic FWM process, the mode-coupling equations can be written as [S2]

$$\frac{dA_1}{dz} = \frac{in_2\omega_1}{c}[(f_{11}|A_1|^2 + 2\sum_{k\neq 1}f_{1k}|A_k|^2)A_1 + 2f_{1234}A_2^*A_3A_4 e^{i\Delta kz}] \tag{S13}$$

$$\frac{dA_2}{dz} = \frac{in_2\omega_2}{c}[(f_{22}|A_2|^2 + 2\sum_{k\neq 2}f_{2k}|A_k|^2)A_2 + 2f_{2134}A_1^*A_3A_4e^{i\Delta kz}] \tag{S14}$$

$$\frac{dA_3}{dz} = \frac{in_2\omega_3}{c}[(f_{33}|A_3|^2 + 2\sum_{k\neq 3}f_{3k}|A_k|^2)A_3 + 2f_{3412}A_4^*A_1A_2e^{-i\Delta kz}] \tag{S15}$$

$$\frac{dA_4}{dz} = \frac{in_2\omega_4}{c}[(f_{44}|A_4|^2 + 2\sum_{k\neq 4}f_{4k}|A_k|^2)A_4 + 2f_{4312}A_3^*A_1A_2e^{-i\Delta kz}] \tag{S16}$$

where $\omega_j/A_j$ ($j = 1, 2, 3, 4$) refers to the angular frequency/amplitude of the pump, signal, idler, and conjugate waves, respectively. The $\Delta\beta$ refers to the effective propagation constant mismatch of the interacting waves. The $n_2$ refers to the nonlinear refractive index in the Kerr process, and $f_{ij}/f_{ijkl}$ ($i, j, k, l = 1, 2, 3, 4$) refers to two-wave overlap integral and four-wave overlap integral, which took the following forms

$$f_{ij} = \frac{<|F_i|^2|F_j|^2>}{<|F_i|^2><|F_j|^2>} \tag{S17}$$

$$f_{ijkl} = \frac{<F_i^*F_j^*F_kF_l>}{[<|F_i|^2><|F_j|^2><|F_k|^2><|F_l|^2>]^{1/2}} \tag{S18}$$

where $f_j$ ($j = 1, 2, 3,$ and $4$) refers to the modal distribution of the pump, signal, idler, and conjugate waves, respectively, and the operator $<\cdot>$ refers to the amplitude integration in the waveguide cross-section.

Similarly, the equations can only be accurately solved using numerical methods and can be expressed as an analytic solution under some simplifying assumptions. If no conjugate wave was injected into the beginning of the waveguide, the output power of the idler wave can be expressed as

$$P_3(z) = P_3(0)[1 + (1 + \frac{\Delta\beta_{FWM}^2}{4g^2})\sinh^2(gz)] \tag{S19}$$

where g denotes the parametric gain in the intrinsic FWM process, and can be expressed as

$$g = \sqrt{(\gamma P_0 r)^2 - (\frac{\Delta\beta_{FWM}}{2})^2} \tag{S20}$$

where $P_0$ denotes the sum of pump power $P_1$ and signal power $P_2$

$$P_0 = P_1 + P_2 \tag{S21}$$

and r denotes the power parameter

$$r = \frac{2\sqrt{P_1P_2}}{P_1 + P_2} \tag{S22}$$

Moreover, $\Delta\beta_{FWM}$ denotes effective phase mismatch in the intrinsic FWM process, which includes linear and nonlinear phase mismatch induced by self-phase modulation (SPM) and cross-phase modulation (XPM). It can be expressed as

$$\Delta\beta_{FWM} = 2\pi\left(\frac{n_{eff,1}}{\lambda_1} + \frac{n_{eff,2}}{\lambda_2} - \frac{n_{eff,3}}{\lambda_3} - \frac{n_{eff,4}}{\lambda_4}\right) + 2\gamma P_0 \tag{S23}$$

$\gamma$ denotes a nonlinear parameter in the intrinsic FWM process and can be expressed as

$$\gamma = \frac{n_2 \omega}{c A_{eff}} \tag{S24}$$

where $n_2$ refers to the nonlinear refractive index of lithium niobate and $A_{eff}$ refers to the effective mode area. It can be approximately expressed as the reciprocal of the modal overlap integral:

$$A_{eff} \approx \frac{1}{f_{ij}} \approx \frac{1}{f_{ijkl}} \tag{S25}$$

Considering zero effective phase mismatch in the intrinsic FWM process, the output power of the idler wave can be expressed as

$$P_3'(z) = P_3(0)[1 + \sinh^2(\frac{2n_2\omega}{cA_{eff}}\sqrt{P_1 P_2} \cdot z)] \tag{S26}$$

Under low idler input power, the conversion efficiency of FWM can be expressed as:

$$\eta_{FWM} = \frac{P_3(z)}{P_2} \approx \frac{P_3'(z)}{P_2} \approx \frac{P_3(0)\sinh^2(\frac{2n_2\omega}{cA_{eff}}\sqrt{P_1 P_2} \cdot z)}{P_2} \tag{S27}$$

Considering cascaded SHG and DFG process as an effective FWM process, we can derive an effective nonlinear refractive index $n_{2,eff}$ from the conversion efficiency expressions:

$$n_{2,eff} = \frac{cA_{eff}}{2\omega z\sqrt{P_1 P_2}}\operatorname{arcsinh}(\sqrt{\frac{\eta P_2}{P_3(0)}}) \tag{S28}$$